\documentclass{osa-article}

\journal{osac}
\articletype{Research Article}
 \usepackage{color}
 \usepackage{float}
 \usepackage{cite}
 \usepackage[normalem]{ulem}
 \usepackage{booktabs}
 \usepackage{amsmath}
 \usepackage{ulem}
 \usepackage{fancyhdr}
\DeclareMathAlphabet\mathbfcal{OMS}{cmsy}{b}{n}
 
 \def\rp{{\bf r}_\perp} \def\cE{\mathbfcal{ E}}  
 \def\sigse{\sigma_s^{\rm(e)}} \def\sigsa{\sigma_s^{\rm(a)}}
 \def\sigpe{\sigma_p^{\rm(e)}} \def\sigpa{\sigma_p^{\rm(a)}}

\begin{document}

\title{Accurate and Efficient Modeling of the Transverse Mode Instability in High Energy Laser Amplifiers}

\author{C. R. Menyuk,\authormark{1,2,*} J. T. Young,\authormark{3} J. Hu,\authormark{3} A. J. Goers,\authormark{1} D. M. Brown,\authormark{1} and M. L. Dennis\authormark{1}}

\address{\authormark{1}Johns Hopkins Applied Physics Laboratory, 11100, Johns Hopkins Road,Laurel, MD 20723\\
\authormark{2}Permanent address:  University of Maryland Baltimore County, 1000 Hilltop Circle, Baltimore, MD 21270\\
\authormark{3}Baylor University, One Bear Place, Waco, TX, 76798}

\email{\authormark{*}menyuk@umbc.edu} %% email address is required



\begin{abstract}
We study the transverse mode instability (TMI) in
the limit where a single higher-order mode (HOM) is present.  We demonstrate
that when the beat length between the fundamental mode and the HOM is small compared to the
length scales on which the pump amplitude and the optical mode amplitudes
vary, TMI is a three-wave mixing process in which the two optical modes
beat with the phase-matched component of the index of refraction that is
induced by the thermal grating.  This limit is the usual limit in
applications, and in this limit TMI is identified as a stimulated thermal Rayleigh 
scattering (STRS) process.  We demonstrate that a phase-matched model that is
based on the three-wave mixing equations can have a large computational
advantage over current coupled mode methods that must use longitudinal
step sizes that are small compared to the beat length.  
\end{abstract}

%%%%%%%%%%%%%%%%%%%%%%%%%%  body  %%%%%%%%%%%%%%%%%%%%%%%%%%
\section{Introduction}
Ytterbium-doped fiber amplifiers that produce kilowatt output powers have 
been developed in the past decade \cite{1_Zervas_JSTQE_2014_219,2LvL_Richardson_JOSAB2010,2_Robin_OL_2014_666,3_Dong_JSTQS_2016_4900207,4_Naderi_OL_2016_3964,5_Gu_CLEO_2017,6_Chen_LP_2019_075103}.  However, the thermal or 
transverse
mode instability (TMI) has become a major barrier to achieving even
higher output powers \cite{Jauregui_AOP_2020_429,Jauregui_NP_2013_861,7_Zervas_OE_2019_19019}.  Despite almost a decade of work since the
original observation of TMI in fiber amplifiers by Jauregui et al. \cite{8_Jauregui_OE_2011_3258} and Eidam et al. \cite{9_Eidam_OE_2011_13218},
it still remains only partially understood, and computationally-efficient
methods that are sufficiently accurate for amplifier design have been
lacking.  It was recognized early by Smith and Smith \cite{10_Smith_OE_2011_10180} that the
instability could be explained by a thermal grating that is induced by the
beating of the fundamental mode of the optical fiber with a higher-order
mode at a slightly lower frequency and the quantum defect heating that ensues. 

Subsequent work by Jauregui
et al. \cite{11_Jauregui_OE_2012_12912}, Dong \cite{12_Dong_OE_2013_2642}, and Smith and Smith \cite{13_Smith_JSTQS_2014_3000112} identified the instability
as a stimulated thermal Rayleigh scattering (STRS) process.  In particular,
Dong \cite{12_Dong_OE_2013_2642} developed a three-wave mixing model that is analogous to models
of the Brillouin instability due to stimulated Brillouin scattering (SBS). This identification has remained somewhat controversial, although
Kong et al. \cite{14_Kong_Optica_2016_975} directly observed the STRS process in a fiber amplifier.
Ward et al. \cite{15_Ward_OE_2012_11407} and Naderi et al. \cite{16_Naderi_OE_2013_16111} developed a model of TMI based on
a coupled mode method that makes no reference to three-wave
interactions. The complexity of TMI has obscured its identification as an STRS-driven, three-wave process. The conditions that are required to treat TMI as
a three-wave instability have not been elucidated.

In this work, we demonstrate that the key requirement is that the beat
length $L_{\rm B} = 2\pi$/$\Delta\beta$ must be small compared to any other longitudinal
scale lengths, where $\Delta\beta$ is the difference between the wavenumber
of the fundamental mode and any higher-order modes (HOMs) at the same frequency.
This condition usually applies in practice.  In this limit, we derive the
three-wave equations that govern TMI\null.  Terms that are not
phase-matched are neglected.  This approach is similar to Dong's 
approach \cite{12_Dong_OE_2013_2642}, but is more general.

We further demonstrate that this approach has a large computational
advantage.  Approaches that use the coupled mode method, like 
the approach of Naderi et al. \cite{16_Naderi_OE_2013_16111}, must use longitudinal steps 
in their computations that are 
small compared to the beat length.  As a result, Naderi et al. \cite{16_Naderi_OE_2013_16111} limited 
their study to a high-gain amplifier with a length of 1.6 m, which is 
substantially shorter than typical ytterbium-doped fiber amplifiers.  
Approaches that use the finite-element method to calculate the optical mode 
profile on each longitudinal step like that of Ward \cite{17_Ward_OE_2013_12053} must take steps 
that are small compared to the optical wavelength and typically require 
large computational resources.  Our approach is only limited by the 
longitudinal scale 
lengths over which the amplitudes of the optical modes, the thermal mode, 
and the pump mode change.  These lengths are typically far larger than the
beat length. For the examples that we consider in this paper, which
correspond to realistic Yb$^{3+}$-doped fiber amplifiers, the 
computational speedup is more than a factor of 100.

\section{Theoretical Model}

We may write the index of refraction as $n(\rp,z,t)=n_0(\rp) + \Delta 
n(\rp,z,t)$, where $n_0(\rp)$ is the unperturbed index of refraction,
and we set $\Delta n\ll n_0$, which is always the case 
($\Delta n/n_0 \sim 10^{-5}$).
We will use the slowly varying envelope approximation, which is an 
excellent approximation in this system due to the large discrepancy between 
the wavelength and the next-smallest scale, which is the intermodal beat
length $L_{\rm B}$.  We will also assume that time derivatives of the
index of refraction can be ignored when calculating the inter-modal
coupling.  That is again an excellent approximation, given the large
disparity between the light period and the time scale on which either
the gain changes (microseconds) or the temperature changes (milliseconds).
We assume as well that the only coupling is between modes that are
propagating in the forward direction.
 We start with the expression from coupled mode theory for two coupled modes \cite{19_Marcuse_TheoryofDielectricWaveguides_1991_Ch3,20_Snyder_OpticalWaveguideTheory_1983_Ch31},
\begin{align}
	\label{eqno1}
	\begin{split}
 		\frac{dA_0}{dz} = \frac{i\omega^2}{\beta c^2}
        \int d^2\rp n_0\Delta n\left[|\cE_0|^2 A_0 + \cE_0^*\cdot\cE_1
        \exp(-i\Delta\beta z)A_1\right],
        \\         
        \frac{d A_1}{dz} = \frac{i\omega^2}{\beta c^2}
        \int d^2\rp n_0\Delta n\left[|\cE_1|^2 A_1 + \cE_1^*\cdot\cE_0
        \exp(i\Delta\beta z)A_0\right],
	\end{split}         
\end{align}
where $\cE_0$ and $\cE_1$ are the normalized transverse mode profiles for the fundamental mode and the HOM, while $A_0$ and $A_1$ are the corresponding amplitudes. We have set $\beta \equiv
	\beta_0 \approx \beta_1$ since $\Delta\beta = \beta_0 - \beta_1 \ll \beta_{0,1}$. Equation (1) is valid when only a single HOM is present or when the amplitudes of the other HOMs are small compared to $A_0$ and $A_1$.

To solve Eq.~(\ref{eqno1}) at any point in time $t$, we must find
$\Delta n({\bf r}_\perp,z,t)$.  Due to the factors $\exp(\pm i\Delta\beta z)$ that appear in Eq.~(\ref{eqno1}), it is necessary to determine $\Delta n({\bf r}_\perp,
z,t)$ with a computational resolution $\Delta z$ that is small compared to
the beat length $L_{\rm B} = 2\pi/\Delta\beta$ even though the field
amplitudes vary slowly compared to this length.  Since we must determine
the transverse dependence of $\Delta n$ as well, this constraint is
computationally demanding.  We can bypass this difficulty by replacing the
field $\Delta n({\bf r}_\perp,z,t)$ with three fields $n_0(\rp,z,t)$,
$n_+(\rp,z,t)$, and $n_-(\rp,z,t)$, which are defined so that
\begin{align}
\label{eqno2}
\Delta n = \Delta n_0 + \frac{1}{2}[\Delta n_+\exp(i\Delta\beta z) 
+\Delta n_-\exp(-i\Delta\beta z),         
\end{align}
and all three fields vary slowly compared to the beat length $L_{\rm B}$.
When we substitute Eq.~(\ref{eqno2}) into Eq.~(\ref{eqno1}) and keep only the phase-matched terms,
we obtain
\begin{align}
\label{eqno3}
\begin{split}
\frac{d A_0}{dz} = \frac{i\omega^2}{\beta c^2}
\int d^2\rp	n_0\left[|\cE_0|^2\Delta n_0 A_0 + \frac{1}{2}\cE_0^*\cdot\cE_1\Delta n_+ A_1\right],
\\         
\frac{d A_1}{dz} = \frac{i\omega^2}{\beta c^2}
\int d^2\rp	n_0\left[|\cE_1|^2\Delta n_0 A_1 + \frac{1}{2}\cE_1^*\cdot\cE_0\Delta n_- A_0\right].
\end{split}         
\end{align}

The terms that are not phase-matched are rapidly oscillating and do not 
contribute significantly to the integrals.  All terms in Eq.~(\ref{eqno3}) vary
slowly in $z$. It then becomes possible to integrate Eq.~(\ref{eqno3}) with
no loss of accuracy while using a
resolution in $z$ that is far larger than is necessary with Eq.~(\ref{eqno1}).

The form of Eq.~(\ref{eqno3}) makes clear that in the limit where Eq.~(\ref{eqno3}) is valid,
TMI is effectively a three-wave process in which two optical fields combine
with a density field.  The condition for Eq.~(\ref{eqno3}) to be valid is that all
terms must vary slowly compared to the beat length.  This condition is
almost always met in practice.  Since the density field is thermally driven,
this process is a stimulated Rayleigh scattering process. We discuss this
identification in more detail in the Appendix.

\begin{figure}	
	\centering\includegraphics[width=10cm]{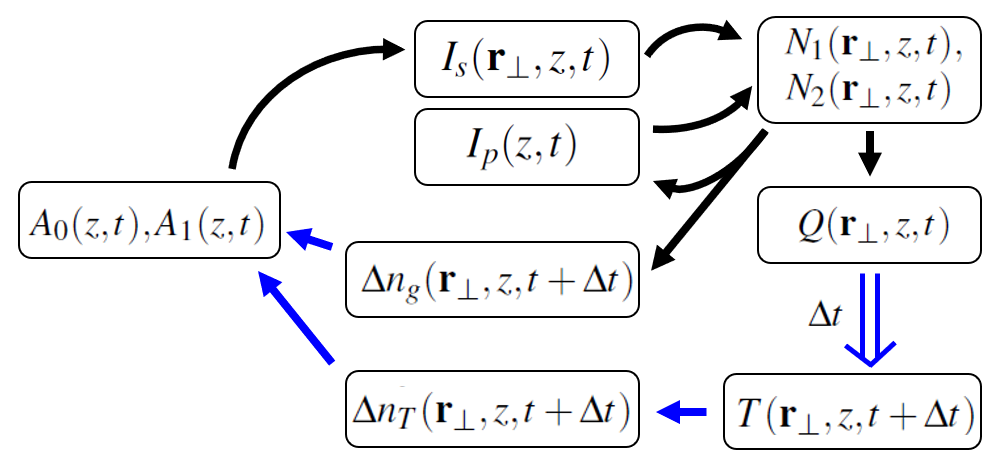}
	\caption{Schematic flow diagram of the computational procedure. The time step occurs when $T(t+\Delta t)$ is calculated using $Q(t)$.}
	\label{Fig0}
\end{figure}

The procedure that we use to obtain $\Delta n_0$, $\Delta n_+$, and $\Delta n_-$ 
in terms of the field and as a function of time is analogous to the
procedure that is used by Naderi et al.~\cite{16_Naderi_OE_2013_16111} that is illustrated in Fig.~\ref{Fig0}.

We begin by writing the signal intensity $I_s(\rp,z,t)$ as
\begin{equation}
\label{eqno4}
I_s(\rp,z,t) = I_{s0} +\frac{1}{2}\left[I_{s+}\exp(i\Delta\beta z) + I_{s-}
\exp(-i\Delta\beta z)\right],
\end{equation}
where
\begin{equation}
\label{eqno5}
I_{s0} = \frac{\beta}{2\mu_0\omega}\left(|\cE_0|^2|A_0|^2 + 
|\cE_1|^2|A_1|^2\right), \qquad I_{s+} = I_{s-}^* =
\frac{\beta}{\mu_0\omega}\cE_1^*\cdot\cE_0 A_1^* A_0,
\end{equation}
all vary slowly compared to $L_B$.  The behavior of a Yb-doped fiber
amplifier is accurately described as a two-level system at realistic
power levels \cite{2LvL_Richardson_JOSAB2010}.  Given $I_s(\rp,z,t)$ and the pump intensity $I_p(z,t)$, we next compute the upper state density $N_2(\rp,z,t)$
of the Yb ions, which is given by
\begin{equation}
\label{eqno6}
N_2 = {\sigpa (I_p/\hbar\omega_p) + \sigsa (I_s/\hbar\omega) \over \left[\sigpa + \sigpe\right](I_p/\hbar\omega_p)
	+\left[\sigsa + \sigse\right] (I_s/\hbar\omega) + 1/\tau}N_0,
\end{equation}
where $\omega$ and $\omega_p$ are the signal and pump angular frequencies, respectively, $\tau$ is the spontaneous decay time of the upper
level, $N_0$ is total density of Yb ions, and $\sigpa$, $\sigpe$, $\sigsa$,
and $\sigse$ are the absorption and emission cross-sections at the pump
and signal frequencies.  Because $I_s$ appears in both the numerator and
denominator of Eq.~(\ref{eqno6}), the density $N_2$ will have higher harmonics that
are proportional to $\exp(\pm im\Delta\beta)$ with $m>1$.  
We will truncate this expression since the harmonics with $m >
1$ are not phase-matched. 
Explicitly, we find that
Eq.~(\ref{eqno6}) has the form
\begin{equation}
N_2 = {A + B\cos\theta\over C + D\cos\theta}N_0,
\label{eqno6b}
\end{equation}
where $\theta=\Delta\beta z + \phi$ and $\phi=\angle(A_1^*A_0\cE_1^*\cdot\cE_0)
$.
We have
\begin{align}
\label{eqno7}
\begin{split}
\hskip2cm A=\sigpa{I_p\over\hbar\omega_p} +\sigsa{I_{s0} 
	\over\hbar\omega}, \qquad B= \sigsa{|I_{s+}|\over\hbar\omega},
\\
C=\left[\sigpa + \sigpe\right]{I_p\over\hbar\omega_p}
+ \left[\sigsa + \sigse\right]{I_{s0}\over\hbar\omega} +{1\over\tau},
\\
\qquad D= \left[\sigsa + \sigse\right]{|I_{s+}|\over\hbar\omega}.
\end{split}
\end{align} 
We note that $D<C$ and $B<C$ is a consequence of $I_{s+}<I_{s0}$, which
in turn follows from the Cauchy-Schwartz inequality.  We now write 
\begin{equation}
N_2\simeq N_{20} + {1\over2}\left[ N_{2+}\exp(i\Delta\beta z) + N_{2-}
\exp(-i\Delta\beta z)\right],
\label{eqno8}
\end{equation}
where
\begin{align}
\label{eqno9}
\begin{split}
N_{20} = {A\over C}{1\over r}\left(1 - {BD\over AC}
{1\over 1+r}\right)N_0,
\\
N_{2+} = N_{2-}^* = {A\over C}{2\over r(1+r)}\left({B\over A}
-{D\over C}\right)N_0\exp(i\phi),
\end{split}
\end{align}
with $r=(1-D^2/C^2)^{1/2}$ \cite{23_Gradshteyn_TableofIntegrals_1980_366}. Equation~(\ref{eqno9}) is an exact
truncation, not an expansion.  We will show in Sec.~3 that the contribution
of higher harmonics with $m>1$ are negligible.

The next stages in the procedure are more straightforward. TMI is generated
by the heat deposition due to the quantum defect between the pump and
the signal, which in turn leads to a time-delayed temperature response that
changes the index of refraction.  The temperature response depends linearly
on the heat deposition, which in turn depends linearly on the upper state
density.  From the expression for the heat deposition $Q$,
	\begin{equation}
		\label{eqno10}
		Q = \left(1-{\omega\over\omega_p}\right)
		\left[\sigpa N_0 - \left(\sigpa + \sigpe\right)N_2\right]I_p,
	\end{equation}
we find
	\begin{align}
	\label{eqno11}
	\begin{split}
		Q_0 = \left(1-{\omega\over\omega_p}\right)
		\left[\sigpa N_0 - \left(\sigpa + \sigpe\right)N_{20}\right]I_p,
		\\
		Q_+ =Q_-^* = - \left(1-{\omega\over\omega_p}\right)
		\left(\sigpa + \sigpe\right)N_{2+}I_p, 
	\end{split}
	\end{align}
where $Q = Q_0 + (1/2)[Q_+\exp(i\Delta\beta z) + Q_-\exp(-\Delta\beta z)]$.
Similarly, from the expression for the temperature evolution,
	\begin{equation}
		\label{eqno12}
		\rho C \frac{\partial T}{\partial t} -\kappa\nabla_\perp^2 T = Q,
	\end{equation}
where $\rho$ is the density, $C$ is the heat capacity, and $\kappa$ is the
heat diffusivity, we find
	\begin{equation}
		\label{eqno13}
		\rho C\frac{\partial T_0}{\partial t} - \kappa\nabla_\perp^2 T_0 = Q_0, 
		\qquad \rho C\frac{\partial T_+}{\partial t} - \kappa\nabla_\perp^2 T_+ = Q_+,
	\end{equation}
where $T = T_0 + (1/2)[T_+\exp(i\Delta\beta z) + T_-\exp(-\Delta\beta z)]$ and $T_- =
T_+^*$.  Integrating Eq.~(\ref{eqno12}) over time in the full model and Eq.~(\ref{eqno13}) over 
time in the phase-matched model, we can now obtain $T(\rp,z,t+\Delta t)
-T(\rp,z,t)$.  Since the temperature tends to a constant $T_{\rm room}$ at
large radius, the appropriate boundary conditions for both $T$ and $T_0$ at
large radius are $T_0=T_{\rm room}$, and the appropriate boundary condition
for $T_+$ is $T_+ = 0$.  This integration is where the basic time step 
occurs, as we show schematically in Fig.~\ref{Fig0}, and it is this step that
is computationally time-consuming.

We can now find the change in the index of refraction.  There are two 
contributions to the index of refraction that we must take into account.  
The first contribution is from the temperature change, for which $\Delta n_T 
= (dn/dT)(T-T_{\rm room})$ and $\Delta n_{T0} = (dn/dT)(T_0-T_{\rm room})$,
$\Delta n_{T+} = (dn/dT)T_+$ in the phase-matched model.  The second
contribution is from the gain,
	\begin{equation}
		\label{eqno14}
		g(\rp,z,t) = \left(\sigse + \sigsa\right)N_2(\rp,z,t) - \sigsa 
		N_0(\rp),
	\end{equation}
from which we find
	\begin{equation}
		\label{eqno15}
		\Delta n_g = -i{c\over2\omega}\left[\left(\sigse + \sigsa\right)N_2(\rp,z,t)
		-\sigsa N_0(\rp)\right].
	\end{equation}
It follows that
	\begin{align}
		\label{eqno16}
		\begin{split}
			\Delta n_{g0} = -i{c\over2\omega}\left[\left(\sigse + 
			\sigsa\right)N_{20} -\sigsa N_0\right],
			\\
			\Delta n_{g+} = -\Delta n_{g-}^* = -i{c\over2\omega}
			\left(\sigse + \sigsa\right)N_{2+},
		\end{split}
	\end{align}
where $\Delta n_g  = \Delta n_{g0} + (1/2)[\Delta n_{g+} \exp(i\Delta\beta z)
+ \Delta n_{g-}\exp(-i\Delta\beta z)]$. Although $\Delta n_g$ is purely imaginary, the components $\Delta n_{g+}$
and $\Delta n_{g-}$ are not.  Naderi et al. \cite{16_Naderi_OE_2013_16111} have pointed out that
the corresponding phase shift does not contribute to the instability, but
plays an important role in the energy balance. Finally, we obtain
$\Delta n_0 = \Delta n_{T0} + \Delta n_{g0}$, $\Delta n_+ = \Delta n_{T+}
+ \Delta n_{g+}$, and $\Delta n_- = \Delta n_{T-} + \Delta n_{g-}$.

To complete the model equations, we must obtain the pump intensity 
$I_p(z,t)$.  We use the expression
	\begin{equation}
		\label{eqno17}
		{d I_p\over dz} = \pm\left[\left(\sigpe + \sigpa\right)\overline{N_2} -
		\overline{N_0}\right]I_p,
	\end{equation}
where the overbar indicates that the population densities are averaged
over the cross-section of the pump.
The sign depends on whether the pump is forward- or backward-
propagating.  Eq.~(\ref{eqno17}) becomes
	\begin{equation}
		\label{eqno18}
		{d I_p\over dz} = \pm\left[\left(\sigpe + \sigpa\right)\overline{N_{20}} -
		\overline{N_0}\right]I_p,
	\end{equation}
in the phase-matched model.

\section{Verification, Accuracy, and Timing of the Phase-Matched Model}

In this section, we first verify the phase-matched model,
 [Eqs.~(\ref{eqno3}), (\ref{eqno7}), (\ref{eqno11}), (\ref{eqno13}), (\ref{eqno16}), (\ref{eqno18})]  by comparing its predictions to those of the full model,  [Eqs.~(\ref{eqno1}), (\ref{eqno6}), (\ref{eqno10}), (\ref{eqno12}), (\ref{eqno15}), (\ref{eqno17})].  We will show that agreement is
excellent for a realistic amplifier system similar to the system that
Naderi et al. \cite{16_Naderi_OE_2013_16111} considered, but using a fiber length of 10 m, which is
a typical experimental length \cite{6_Chen_LP_2019_075103,31_Li_PJ_2018_085109}.  We then consider in more detail the
error as a function of the step size $\Delta z$ and show that the
phase-matched model has a significant computational advantage.

\medskip
\subsection{Verification}

\begin{figure}	
	\centering\includegraphics[width=10cm]{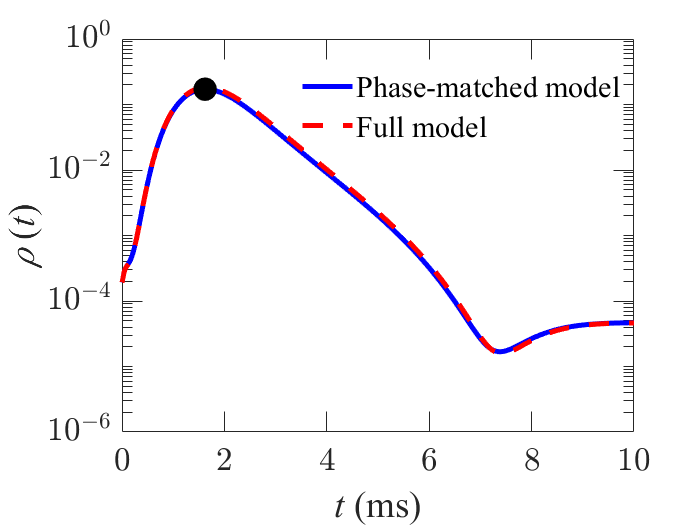}
	\caption{Power ratio $\rho(t)= P_{\rm HOM}(t)/P_{\rm total}(t)$ at the end of the amplifier vs. time {\it t}. The pump power equals 250 W.}
	\label{Fig1}
\end{figure}

\begin{figure}	
	\centering\includegraphics[width=10cm]{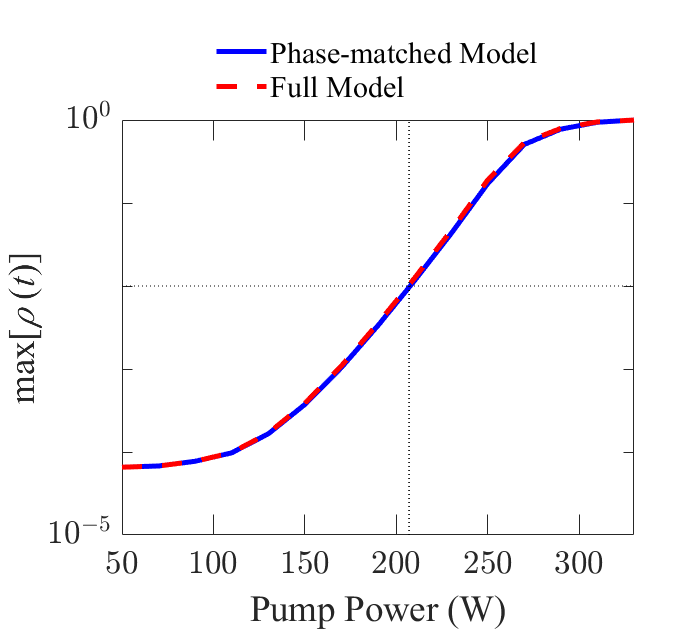}
	\caption{Maximum power ratio $\max[\rho(t)]$ at the amplifier end 
		vs.\ input pump power $P_{\rm pump}$.  The threshold ratio of 1\% occurs
		when $P_{\rm pump}$ = 207 W.}
	\label{Fig2}
\end{figure}

We show the basic set of parameters that we are considering in Table 1.
These parameters are similar to those used in \cite{16_Naderi_OE_2013_16111}, but with a more
realistic amplifier length of 10 m.  We use the alternating-direct-implicit
(ADI) method to integrate the temperature equations, and we used the
Runge-Kutta method to carry out the $z$-integration.  In all the
simulations reported here, we used a $140\times140$~$\mu$m$^2$ square grid
for the transverse integration with a $2\times2$~$\mu$m$^2$ grid spacing.
We verified that using a larger grid size of $160 \times 160 ~\mu {\rm m}^2$ with a smaller grid spacing of $1 \times 1 ~\mu {\rm m}^2$ makes a negligible difference in Figs. \ref{Fig1} and \ref{Fig2}.
We chose the
{\it z}-step sufficiently small so that the relative error is below
1\%. In all the simulations reported here,
we used a noise seed ratio at the entry to the optical fiber of $10^{-4}$ 
between the higher-order mode and the fundamental mode.  We verified that 
using a noise seed ratio of $10^{-3}$ or $10^{-5}$ does not change the 
agreement between the full and phase-matched models that we present in Figs. \ref{Fig1} and \ref{Fig2}.

\begin{table}[b]
	\centering
	\centering{\bf Table 1.  Simulation Parameters}	
	\vspace{0.05 in}
	\centering
	
	\begin{tabular}{@{}|l|l|@{}}
		
		\toprule
		${\it L}_{\rm fiber}$ = 10 m  & ${\it n}_{\rm core}$ = $1.45031$ \\ \midrule
		${\it D}_{\rm core}$ = 50 $\it \mu$m & ${\it N.A.}$ = 0.03 \\ \midrule
		${\it D}_{\rm cladding}$ = 250 $\it \mu$m & $\sigma_p^{(e)}$ = $1.87 \times 10^{-27} ~\rm m^2$  \\ \midrule
		${\it N}_{0}$ = $1 \times 10^{25}  ~\rm m^{-3}$ & $\sigma_p^{(a)}$ = $1.53 \times 10^{-24} ~\rm m^2$ \\ \midrule
		${\it \lambda}_{\rm pump}$ = $977$ nm & $\sigma_s^{(e)}$ = $6 \times 10^{-27} ~\rm m^2$  \\ \midrule
		${\it \lambda}_{\rm signal}$ = $1064$ nm & $\sigma_s^{(a)}$ =  $3.58 \times 10^{-25} ~\rm m^2$ \\ \bottomrule
	\end{tabular}
\end{table}

In Fig. \ref{Fig1}, we show a comparison of the ratio $\rho(t) = P_{\rm HOM}(t)/P_{\rm total}(t)$ of the power in
the higher-order mode $P_{\rm HOM}(t)$ to the total power 
$P_{\rm total}(t)$ at the end of the fiber as a function of time.  In the case that we show
here, the input pump power $P_{\rm pump}$ equals 250 W.  With this pump power, we find that $\rho(t)$ rises to 
a maximum of 17\%, shown as a dot in Fig. \ref{Fig1}, before returning to a value 
that is close to the initial higher-order mode seeding.  We
observe excellent agreement between the full model and the phase-matched
model.  As the pump power increases, we continue to observe excellent
agreement between the two models, although when the pump power is large enough where
both models predict chaotic oscillations, the agreement is qualitative
rather than quantitative.  This behavior is expected since small changes
in the seeding ratio also produce large changes in this limit due to the
butterfly effect \cite{32_Hilborn_ALP_2004_425}.

In Fig. \ref{Fig2}, we show $\max[\rho(t)]$ vs. $P_{\rm pump}$ and a dotted line
that corresponds to a ratio of 1\%.  We observe excellent agreement
between the phase-matched model and the full model.  In this work, we define 
the threshold power as the lowest pump power at which 
$\max[\rho(t)] > 0.01$, i.e., the ratio of $P_{\rm HOM}(t)/P_{\rm total}(t)$
exceeds 1\% at any time.  Beyond this threshold, the beam quality rapidly
degrades \cite{8_Jauregui_OE_2011_3258,9_Eidam_OE_2011_13218,33_Otto_OE_2012_15710,34_Otto_OE_2013_19375}.  This definition of the threshold is consistent with
studies of amplifier limits due to SBS.  In the case considered
here, the threshold power equals 207 W\null.

\begin{figure}	
	\centering\includegraphics[width=10cm]{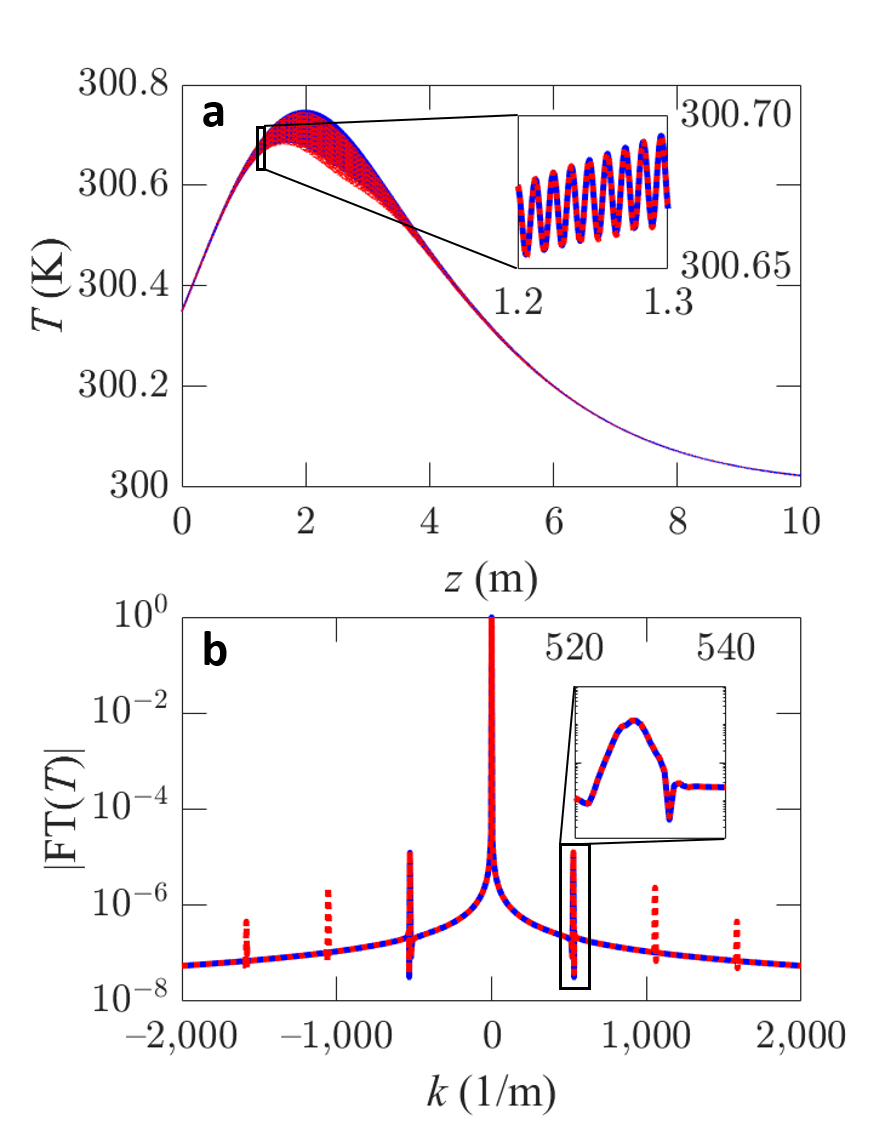}
	\caption{Temperature $T$
		at a point that is 10 $\mu$m from the amplifier center when $t=0.5$ ms.  (a) $T$ vs. $z$.  
		Agreement is excellent between the full model and the phase-matched
		model.  The inset shows that the details of the temperature oscillations
		agree.  (b) $|$FT($T$)$|$ vs.~$k$, where $k$ is the spatial Fourier transform
		variable.  Agreement for the central harmonic and the harmonics at
		$k=\pm\Delta\beta$ is excellent.  The inset shows excellent agreement for
		the harmonic at $k=\Delta\beta$.  However, the harmonics at $k=\pm n\Delta
		\beta$ for $n\ge2$ are not present in the phase-matched model.}
	\label{Fig3}
\end{figure}

In Fig. \ref{Fig3}, we show the temperature as a function of longitudinal position
$z$ and the absolute value of its spatial Fourier transform at a distance of 10~$\mu$m from the 
amplifier center and a time $t=0.5$ ms.  We set $P_{\rm pump} = 450$~W\null.
In Fig. \ref{Fig3}(a), where we show the
temperature as a function of position, we see that the agreement between
the full model and the phase-matched model appears excellent.  In the
inset, where we show the spatial oscillations, the two models appear
indistinguishable.  However, the subtle difference is visible in Fig. \ref{Fig3}(b),
where we show the absolute value of the spatial Fourier transform, $|\hbox{FT}(T)|=|\int T(z)
\exp(ikz)\,dz|$.  Agreement is excellent
for the central harmonic, as well as the two surrounding harmonics 
which are located at $k=\pm\Delta\beta = \pm 528$ m$^{-1}$.  However, the
phase-matched model has no contribution from the harmonics at $\pm 
n\Delta\beta$, where $n\ge2$.  It is precisely these higher harmonics
that we are neglecting.

\medskip
\subsection{Accuracy and Timing}

In the phase-matched model, the number of 
dependent variables is almost twice as large as in the full model.
In particular, it is necessary to solve the temperature equation, Eq. (\ref{eqno13}), 
for both $T_0$ and $T_+$ instead of just solving the temperature equation,
Eq. (\ref{eqno12}), for the single temperature $T$.  As a result, we have found that
the computational load on each $z$-step increases by roughly a factor of
two.  However, it is possible to take
significantly larger steps, leading to a large computational advantage. 

\begin{figure}	
	\centering\includegraphics[width=10cm]{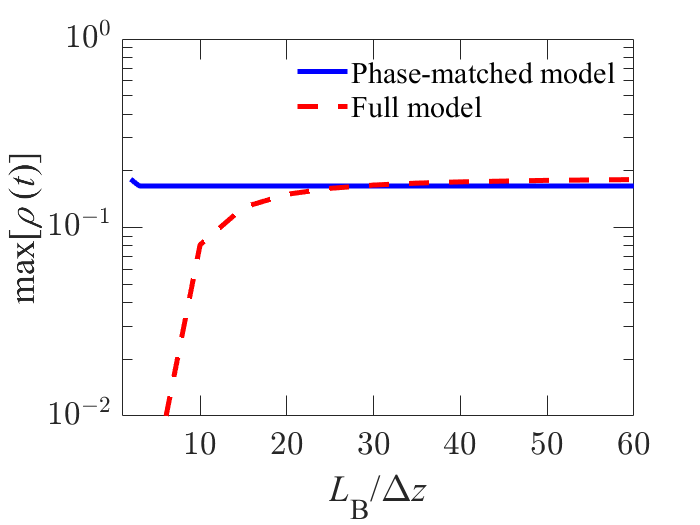}
	\caption{Convergence of the
		maximum power ratio at the amplifier end as $L_B/\Delta z$ increases.
		We set $P_{\rm pump} = 250$ W.}
	\label{Fig4}
\end{figure}

In Fig. \ref{Fig4}, we show $\max[\rho(t)]$ for both the full model and phase-matched
model as a function of $L_{\rm B}/\Delta z$ when $P_{\rm pump} = 250$~W,
so that the pump power is slightly above threshold.  As expected, the
full model requires an $L_{\rm B}/\Delta z > 60$ to converge, while the phase-matched model appears to have converged in this
case when $L_{\rm B}/\Delta z\simeq2$.  

In order to quantify the convergence, we define the relative error,
$\epsilon$, as the difference between our computation
at a given $\Delta z$ and a four-point Richardson extrapolation \cite{35_W_H_Press_Rich_Extrap}.  For
the full model, we used $L_{\rm B}/\Delta z = 80$, 40, 20, and 10
for the extrapolation.  For the phase-matched model, we used $L_{\rm B}/
\Delta z = 40$, 20, 10, and 5.

\begin{figure}	
	\centering\includegraphics[width=10cm]{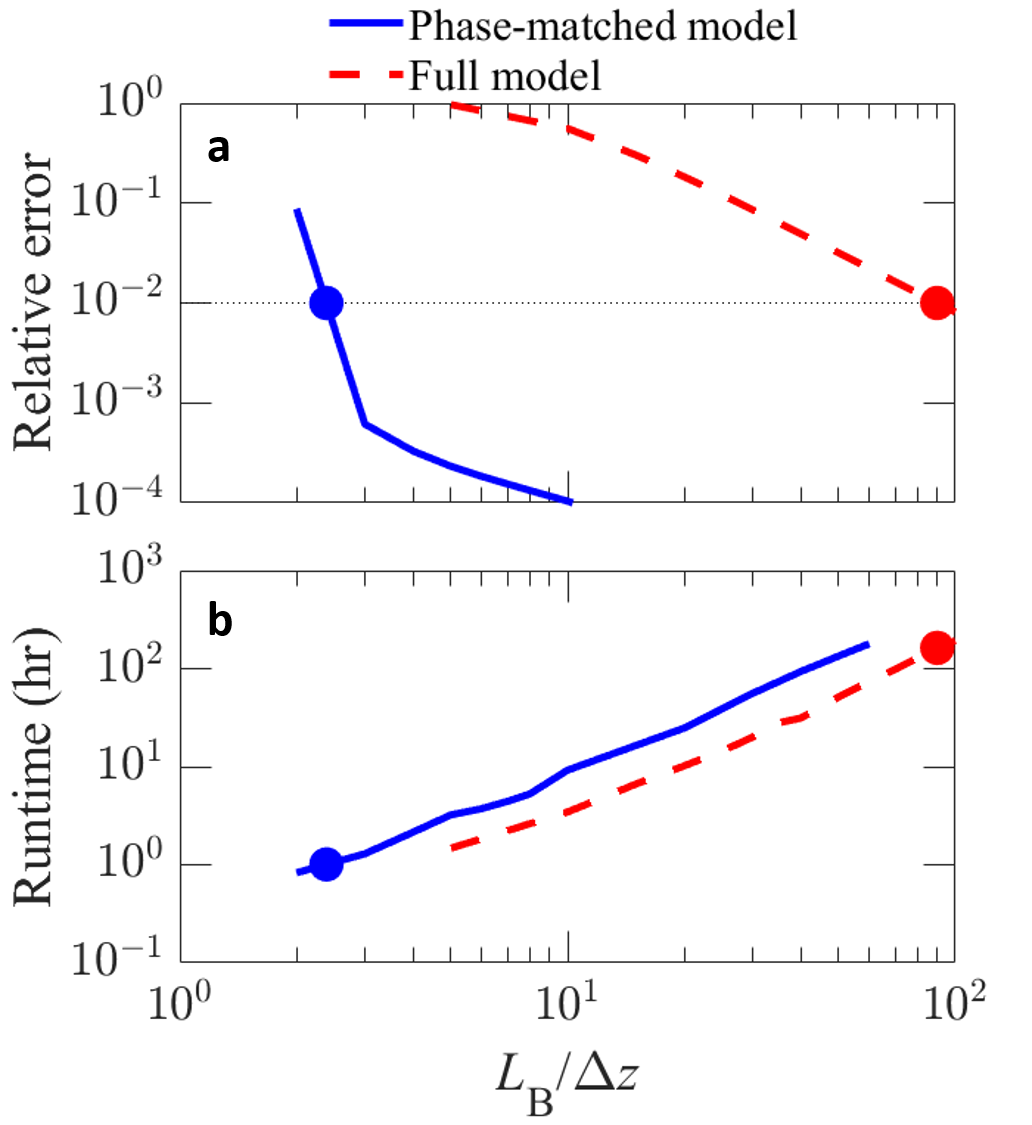}
	\caption{(a) Relative error vs.~$L_B/\Delta z$ for the cases shown in
		Fig. 4. (b) Runtime vs. $L_{\rm B}/\Delta z$.  The runtime for the phase-matched model is approximately twice as long as the runtime for the full model. Dots indicate the corresponding runtime with a relative error of $1\%$, as shown in Fig. 5.}
	\label{Fig5}
\end{figure}

In Fig. \ref{Fig5}(a), we show the relative error as a function of $L_{\rm B}/\Delta z$.
We see that achieving a relative error of 1\% with the full model
requires $L_{\rm B}/\Delta z \simeq90$, while the same relative error can be
obtained with the phase-matched model when $L_{\rm B}/\Delta z \simeq 2$.
Figure \ref{Fig5}(a) also illustrates that the full model is second-order accurate
in $\Delta z$, so that the relative error decreases proportional to
$(\Delta z)^{-2}$.  This result is consistent with the result of Naderi
et al. \cite{16_Naderi_OE_2013_16111}, but may be surprising since our integration in $z$ is done
using the Runge-Kutta method.  This result indicates that the global
error is dominated by the accumulated error in computing the index of
refraction.  The variation of the relative error in the phase-matched
model is more complex since it depends on the rate at which all the
dependent variables change as a function of $z$.  A complete error
analysis is beyond the scope of this paper, but Fig. \ref{Fig5}(a) indicates that
it decreases rapidly as $\Delta z$ increases until it has become quite
small.

In Fig. \ref{Fig5}(b), we show the runtime of both the full model and the phase-matched model using 16 cores in a shared memory system that consists of dual Intel E5-2695 V4 processors. We observe that the runtime for the phase-matched model is approximately twice as the runtime for the full model, which is consistent with the greater computational
load per step in the phase-matched model.  When we compare the runtime 
corresponding to a relative error of 1\%, shown with dots, we found a runtime of 163 hours for the full model and 1.17 hours for the phase-matched
model, indicating that the phase-matched model runs a factor of 139 faster
in this case. While we obtain a speedup of 139 in our study, this number will vary with different simulation parameters. Nevertheless, the advantage of using phase-matched model is clear.

\section{Conclusions and Discussion}
In this work, we derived the three-wave mixing equations that govern
TMI in the limit where a single higher-order mode is present, and the 
longitudinal rate of change of all quantities is slow compared to the beat 
length.  This limit normally applies in practice.  In this limit, TMI
can be identified as an STRS process, and we reviewed the theoretical
justification for this identification in the appendix, where we also discussed similarities
and differences between STRS and SBS in the Appendix. There, we verified the accuracy
of the phase-matched model in a Yb$^{3+}$-doped fiber amplifier with a relatively
simple step index profile.  The amplifier that we considered is like that
of Naderi et al. \cite{16_Naderi_OE_2013_16111}, but has a more realistic 10-m length.  We
demonstrated that this model reproduces the nonlinear saturation of the
higher-order mode and the instability threshold that are predicted by
the full model.  We demonstrated a computational speedup that is more than
a factor of 100.

We derived the three-wave mixing equations in the case that a single
higher-order mode is present, but we expect this result to be more
broadly applicable when several higher-order modes are present.  In the
linear limit below threshold, each higher-order mode will only interact
with the fundamental mode.  In that case, the three-wave mixing equations
can be extended by adding a new set of equations for the index of
refraction, Yb$^{3+}$ population density, heat, temperature, and optical mode
amplitude for each of the higher-order modes. The computational
complexity scales proportional to $M$, where $M$ is the number of modes.
More generally, we anticipate that the three-wave mixing equations can
be extended to include a coupling between all the modes as long as none
of the beat lengths between any of mode pairs becomes large enough to
be comparable to the scale length on which any of the amplitudes change.
However, the computational complexity grows proportional to $M^2$, and
higher-order nonlinear interactions with a slowly varying amplitude
could invalidate this approach.

\begin{backmatter}

\bmsection{Acknowledgments}
Work at the Johns Hopkins Applied Physics Laboratory (JHU-APL) and at 
Baylor University was supported by JHU-APL Internal Research and 
Development Funds. The authors would like to thank R. A. Lane, J. R. Grosek, and E. J. Bochove at the Air Force Research Laboratory in Kirtland, NM for helpful discussions.

\bmsection{Disclosures}
\noindent The authors declare no conflicts of interest.

\bmsection{Appendix}

{\it TMI as an STRS process}

The identification of TMI as an STRS process has 
remained somewhat controversial due to the complexity of TMI\null.  
Here, we will briefly argue in favor of this identification in the limit where the phase-matched model holds. We then point
out some of the similarities and differences with the instability due
to stimulated Brillouin scattering (SBS), which is another important
effect limiting the performance of high-energy fiber laser amplifiers \cite{24_Dawson_OE_2008_13240}.

Rayleigh scattering is commonly observed as a spontaneous process.  It
is well known as the reason the sky is blue \cite{25_See_PrinciplesofOptics_Ch14} and imposes a fundamental
loss limit on optical fiber transmission \cite{26_See_OpticalFiberComm_Ch3}.  It also imposes a
fundamental limit on fiber interferometers and hence on opto-electronic
oscillators \cite{27_Fleyer_OE_2015_25635}.

Observation of STRS has proved more
elusive, particularly in optical fibers.  Zhu et al. \cite{28_Zhu_OE_2010_22958} reported
an observation of STRS in 2010, and Kong et al. \cite{14_Kong_Optica_2016_975} reported an
observation of STRS in 2016.  It is difficult to observe directly, and
another observation that was reported in 2012 \cite{29_Okusaga_OL_2012_683} was later shown to
be incorrect \cite{30_Okusaga_OL_2013_549}.

STRS and SBS can be treated together theoretically because both are due to 
density fluctuations \cite{31_Li_PJ_2018_085109}.  Rayleigh scattering is driven by isobaric 
processes, while Brillouin scattering is driven by isentropic processes. 
Both are three-wave scattering processes in which two optical fields
couple to density fluctuations. When Eq.~(\ref{eqno3}) holds, it is evident
that TMI can be treated as a three-wave scattering process
in which two optical modes couple to density fluctuations and that this
process is isobaric.  Hence, it is reasonable to identify TMI as an
STRS-driven process.

While both STRS and SBS are three-wave processes in which two optical
modes couple to density fluctuations, there are important differences---particularly in optical fibers.
Rayleigh scattering is often referred to as an inelastic process, but
that is almost never strictly true.  Energy and momentum conservation
implies that there is typically a small frequency offset.  In the case
of TMI this offset is quite small---on the order of a few kilohertz 
\cite{10_Smith_OE_2011_10180,11_Jauregui_OE_2012_12912,12_Dong_OE_2013_2642,13_Smith_JSTQS_2014_3000112}. This offset plays a critical role in driving the instability, but it lies well within the linewidth of the optical
modes, which is typically on the order of 100 MHz.  While
there is a significant difference between the wavenumbers of the
fundamental and higher-order modes, this difference is small compared
to the wavenumber of both modes ($\Delta\beta / \beta \sim 10^{-5}$).
Both modes propagate in the same direction, but have different mode
profiles.  By contrast, the two optical modes that become unstable due
to SBS are both fundamental modes, but they propagate in opposite
directions.  As a consequence, the wavenumber offset equals twice the
wavenumber of each of the optical modes.  The frequency offset, which is given
by (the acoustic velocity)$\times$(twice the wavenumber of the
optical modes), is  of 10--20 GHz and much larger
than the linewidth of the optical modes.  These differences can be traced
to the fundamental physical difference between pressure fluctuations,
which propagate, and entropy fluctuations, which do not.

\end{backmatter}
%%%%%%%%%%%%%%%%%%%%%%% References %%%%%%%%%%%%%%%%%%%%%%%%%

\end{document}